\documentclass[twocolumn,prl,aps]{revtex4}
\usepackage{amssymb}
\usepackage{epsfig}

\begin{document}

\title{Complementarity and  diversity in
a soluble model ecosystem}

\author{Viviane M. de Oliveira and J.\ F.\ Fontanari }
\affiliation{Instituto de F\'{\i}sica de S\~ao Carlos,
  Universidade de S\~ao Paulo,
  Caixa Postal 369, 13560-970 S\~ao Carlos SP, Brazil}

\begin{abstract}
Complementarity among species with different traits is one of the basic
processes affecting biodiversity, defined as the number
of species in the ecosystem. We present here a soluble
model ecosystem in which the species are characterized by  binary 
traits and their pairwise interactions follow a complementarity principle.
Manipulation of the species composition, and so the
study of its effects on the species diversity, is achieved through the
introduction of a bias parameter favoring one of 
the traits.  Using statistical mechanics tools
we find explict expressions for the allowed values 
of the equilibrium species concentrations in terms of the control parameters of the model.
\end{abstract}

\maketitle

One of the main interests in the study of model ecosystems  is to elucidate the
rules governing the assemblage of ecological communities \cite{May}. In particular, 
understanding the critical role that species composition  
plays in ecosystem processes can provide useful guidelines  to  management efforts 
on behalf of endangered species \cite{Kareiva}. Theory has been prominent
in the study of long-term effects of species composition on ecosystem functioning, 
owing mainly to the difficulty to keep controlled experimental conditions in place for
a long period (for a recent remarkable exception, see \cite{Brown}).
In these studies,  $N$   species  are considered to interact in 
a given community so that their population numbers or biomasses, denoted by 
$x_1, \ldots, x_N$, are determined by the dynamical equations
\begin{equation}\label{geral}
\frac{\partial x_i}{\partial t} =  x_i G \left ( x_i, \sum_{j}^N J_{ij} x_j  \right )
\end{equation}
where $G (~)$ is a species-independent nonlinear function, such as Ricker dynamics
\cite{Murray}, giving the rate of increase of each
population and $J_{ij}$ is the interaction coefficient measuring the effect of species $j$ on
species $i$ \cite{Ives,Lundberg}. As a rule, the number of species (diversity) 
$N$ is taken as the control parameter
or independent variable and the total biomass (productivity) $\sum_i x_i$, viewed as
a measure of  the ecosystem stability, as the 
dynamical variable \cite{Ives,Tilman,Lundberg}. 
However, especially in long-term studies of large 
ecosystems this approach is faulty since  it does not  take into 
account the dynamics of diversity and the species potencial for adaptation
to environmental fluctuations \cite{Loreau}.

In this contribution we consider
an alternative approach in which the diversity is given by the number of surviving 
species in the ecosystem at equilibrium, being thus a dynamical variable that,
ultimately, 
depends   on the nature of the  species composing the ecosystem. 
In particular, we assume that
the fraction or concentration of individuals of species  $i$ in the ecosystem,
$x_i \in [0,\infty)$, is determined by the nonlinear system of equations, so-called
replicator equations \cite{Hofbauer},
\begin{equation}\label{eq_rep0}
\frac{\partial x_i}{\partial t} =  x_i \left ( {\mathcal{F}}_i -
\frac{\phi}{N} \right ) 
\end{equation}
where ${\mathcal{F}}_i = \sum_j J_{ij} x_j $ 
can be identified with the fitness of species $i$ and 
the term $ \phi = \sum_i x_i {\mathcal{F}}_i $ ensures that   
\begin{equation}\label{eq_const}
\sum_{i=1}^N x_i = N 
\end{equation}
for all times. This constraint enforces an effective 
competition among the species in the ecosystem. 
We note that by dropping
the  $\phi$ term in Eq. (\ref{eq_rep0}) we recover the
classical equation of Lotka-Volterra \cite{Murray}.

In the case of symmetric interactions $J_{ij} = J_{ji}$
the asymptotic regime of Eq. (\ref{eq_rep0}) is simply 
characterized: the dynamics {\it maximizes} the Lyapunov function or fitness
functional 
\begin{equation}\label{eq_functional}
{\mathcal{F}} \left ( \{ x_i \} \right ) = - \sum_{i,j} J_{ij} x_i x_j 
\end{equation}
and so it can be shown that the only stationary states are fixed  points \cite{Hofbauer}.
In this case  $J_{ij} < 0$ corresponds to pairs of cooperating species
whereas $J_{ij} > 0$  to pairs of competing species. The equilibrium regime 
as well as some aspects of the dynamics can become nontrivial, however, in the case 
that the functional ${\mathcal{F}}$ has many local maxima. This occurs 
when the coupling strengths $J_{ij}$  are quenched random variables taking on positive 
and negative values,
as in the model of random replicators put forward  by Diederich 
and Opper \cite{Diederich} (for further development
see \cite{Opper,Oliveira1}). The assumption of random
interactions is a form of taking into account our lack of knowledge 
of how the species actually interact. Moreover, it
 allows  the use of tools of the statistical mechanics of disordered 
systems to fully characterize the equilibrium states, thus making feasible
the study of large ecosystems. In contrast, to keep  numerical accuracy under control,
the traditional approach
based on the numerical solution of Eq. (\ref{geral}) is
restricted to ecosystems composed of typically $N=10$ species.
We note that in the random replicator framework the productivity $\sum_i x_i$ 
is constant, while the diversity varies since a fraction of the
$N$ species may go extinct due to outcompetition. This phenomenon becomes
appreciable, and hence passive of quantitative analysis, for large $N$ only. 

Up to now studies of the random replicator model have considered the
strengths of the interactions between species as independent,
Gaussian distributed random variables \cite{Diederich,Opper,Oliveira1}.
In this contribution we go beyond
this initial stage by introducing some underlying, non-random
structure in the
interspecies interactions. In the spirit of models for molecular recognition
\cite{Farmer,Lancet} we assume that
each species is characterized by a set of $p$  traits, 
$\xi_i^\mu, \mu = 1, \ldots, p$,
and that the resulting interactions between pairs of species depend on these
traits according to a complementarity principle. Specifically,
we assume that the traits $\xi_i^\mu$ are quenched, independent random variables 
that can take on the values
$+1$ and $-1$ with probabilities $(1+a)/2$ and $(1-a)/2$, respectively.
Here $a \in [0,1]$ is a bias towards the trait $+1$, corresponding, e.g., to
an economically  favored  feature of the species. In this sense,
the parameter $a$ may be thought of as a measure of the human impact on the 
species composition of the ecosystem.  
In addition, we
assume that the coupling strength
between species $i$ and $j$ is given by the Hebb rule
\begin{equation}\label{Hebb}
J_{ij}= \frac{1}{2N} \sum_{\mu=1}^p \xi_{i}^{\mu} \xi_{j}^{\mu} ~~~~i \neq j
\end{equation}
that was extensively studied in the eighties within the neural networks context 
\cite{Hopfield,Amit_book}. Clearly, the larger the number of complementary traits
(i.e., $\xi_i^\mu \xi_j^\mu = -1$), the more cooperative the pair of species.
Complementarity among species with different traits has been suggested as
one of the major mechanisms involved in biodiversity. 
In that context,  the term complementarity subsumes all
local deterministic processes
which increase the performance of  communities above that expected from the performance of
individual species grown alone \cite{Loreau}. If community performance
is measured in terms of the fitness functional (\ref{eq_functional}) then the
Hebb coupling between species (\ref{Hebb}) is clearly well suited to model such  processes.

As the final ingredient to define the model,
 we need to specify the self-interactions $J_{ii}$. 
Though for finite $N$ the constraint (\ref{eq_const}) prevents the unbounded
growth of any single species, it becomes inefficacious in the
thermodynamic limit $N \to \infty$ and so an additional mechanism to limit
growth  becomes necessary. The usual procedure is to introduce a competition
term between individuals of a same species by setting  $J_{ii} = u > 0$ for
all $i$ so that  $u$ can be viewed  as a global cooperation 
pressure. In practice, a positive self-interaction is essential to
ensure convergence of the numerical methods used to solve the replicator 
equations (\ref{eq_rep0}).

The maxima of the functional given in Eq. (\ref{eq_functional}) can easily
be obtained within  the statistical mechanics framework in the limit of infinite 
$N$ but finite $p$ (see, e.g., \cite{Amit_book}). 
We begin by defining the free-energy density $f$ as
\begin{equation}\label{f1}
-\beta f= \lim_{N \to \infty} \frac{1}{N}  \ln Z 
\end{equation}
where
\begin{eqnarray}\label{Z0}
Z &=& \int_0^{\infty} \prod_i {\mathrm{d}} x_i ~ \delta( N-\sum_i x_i) \nonumber\\
&& \times \exp \left [
-  \frac{\beta}{2N} \sum_{\mu} \left( \sum_i \xi_i^{\mu} x_i \right)^2 
- \beta u  \sum_i x_i^2 \right] 
\end{eqnarray}
is the partition function and $\beta=1/T$ is the inverse temperature.  Taking the
limit $T \rightarrow 0$ in Eq. (6) ensures that only the states that maximize 
$\mathcal{F}$ will contribute to $Z$.  After some standard algebric manipulations
\cite{Amit_book}, we find
\begin{eqnarray}
-\beta f &=& Q + \frac{\mathbf{m}^2}{2 \beta} + \ln {\left\{ \frac{1}{2} \sqrt{\frac
{\pi}{\beta u}}\right\}} + 
\frac{ \left\langle \left ( Q + \mathbf{m} \cdot \mathbf{\xi} \right )^2 \right \rangle }
{4\beta u} \nonumber \\ 
&& + \left\langle \ln \mbox{erfc} \left( \frac{Q+ \mathbf{m} \cdot \mathbf{\xi}}
{2\sqrt{\beta u}} \right) \right\rangle
\end{eqnarray}
with the notation $\mathbf{z} \cdot \mathbf{y} = \sum_\mu z^\mu y^\mu$ and
where we have invoked the self-averaging property
of the sums $(1/N) \sum_i g (\xi_i^\mu)$, where $g$ is any continuous function,
to replace them by the species-independent averages $\langle g (\xi^\mu) \rangle$.
Here  $\langle \ldots \rangle$ stands for the average taken with the 
probability distribution
\begin{equation}
{\mathcal{P}} \left ( \mathbf{\xi} \right ) = \prod_\mu \left [
\frac{1+a}{2}~\delta \left ( \xi^\mu - 1 \right ) +
\frac{1-a}{2} ~\delta \left ( \xi^\mu + 1 \right ) \right ] .
\end{equation}
The saddle-point parameters $Q$ and $m^\mu$ are given by the
solutions of the  equations $\partial f/\partial Q =0$ and
$\partial f/\partial m^{\mu}=0$. Only $m^\mu$ has a relevant physical
meaning, namely,  it is the average overlap between the equilibrium solutions and
the trait $\xi_i^\mu, i = 1, \ldots, N$,
\begin{equation}
m^\mu = \left\langle \frac{1}{N} \sum_i \xi_i^\mu \langle x_i \rangle_T \right\rangle
\end{equation}
where $\langle \ldots \rangle_T$ 
stands for a thermal average taken with the 
probability distribution
\begin{equation}\label{Gibbs}
{\mathcal{W}}(\{ x_i \})=\frac{1}{Z}\delta (N-\sum_i x_i) 
\exp \left [\beta {\mathcal{F}}(\{ x_i \}) \right ] .
\end{equation}
The next step is to take the zero-temperature limit $\beta \to \infty$.
Nonzero solutions for $Q$ and $m^\mu$  are found in the regime 
$Q+ \mathbf{m} \cdot \mathbf{\xi} < 0$ only, where the $p+1$ saddle-point 
equations take on a particularly simple form, whose only solution is
the symmetric one $m^\mu = m~\forall \mu$. Explicitly, we find
\begin{equation}
\hat{m}=\frac{2ua}{2u + 1 - a^2}
\end{equation}
and
\begin{equation}
\hat{Q} = - 2u \left ( 1 + \frac{p a^2}{2u + 1 - a^2} \right ) 
\end{equation}
with the notation $\hat{Q} = Q/\beta$ and $\hat{m}=m/\beta $.

\begin{figure}[tbc]
\centerline{\epsfig{file=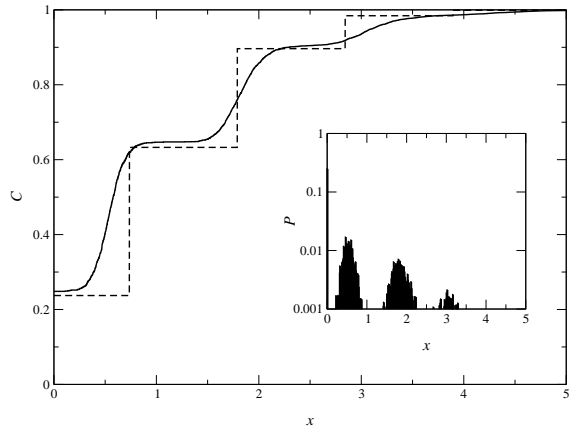,angle=270,width=0.49\textwidth}}
\bigskip
\caption{Cumulative distribution of species for $p=5$, $u=0.1$ and 
$a=0.5$. The  dashed curve is the  prediction
of  Eq. (\ref{cum}) and the solid curve is the result of the numerical 
solution of the replicator equations. The inset shows the histogram
associated to the probability distribution $P(x)$.}  
\label{fig:1}
\end{figure}

The characterization of the ecosystem through the global parameter
${\mathbf m}$ is not very illuminating, and a
better understanding can be achieved by looking directly at the values the
species concentrations $x_i$ can take on. This can be done, for instance, by
calculating explicitly
the cumulative distribution that the concentration of a given
species, say $x_k$, assumes a value smaller than $x$, defined by 
\begin{equation}\label{prob_x}
C_k \left ( x \right )
 =    \lim_{\beta \to \infty} \int_0^\infty \prod_j dx_j ~
\Theta \left ( x - x_k \right )
{\mathcal W} \left ( \{ x_i \} \right )  
\end{equation} 
where $\Theta (x) = 1$ if $ x \geq 0$ and $0$ otherwise, and
${\mathcal W} \left ( \{ x_i \} \right )$ is given by 
Eq.\ (\ref{Gibbs}). Since all species concentrations are equivalent we
can write $C_k \left ( x \right ) = 
C \left ( x \right ) \forall k$ and 
evaluate Eq.\ (\ref{prob_x}) by adding the  field
term $ h \sum_i \Theta \left ( x - x_i \right )$
to Eq.\ (\ref{eq_functional}). Taking the derivatives of the resulting
free-energy with respect to $h$ and then the limit $h \to 0$ yield 
\begin{equation}\label{cum}
C(x)= \sum_{n=0}^p W_n 
\Theta \left ( x - \zeta_n \right ) ,
\end{equation}
where
\begin{equation}
W_n =  \left ( \begin{array}{c} p \\ n \end{array} \right )
\left ( \frac{1 + a}{2} \right )^{n}
\left ( \frac{1 - a}{2} \right )^{p-n}
\end{equation}
and
\begin{equation}\label{zeta}
 \zeta_n = 1 + \frac{p a \left( a + 1 \right ) - 2a n}{2u + 1 - a^2}
  ~~~~~~n=0,1,\ldots,p .
\end{equation}
Hence the species concentrations can take on the values 
$x = \zeta_n$  (provided that $\zeta_n > 0$) only,  and the fraction of species
with concentration $\zeta_n$ is given by $W_n$.
Here the integer variable $n$ yields the total number of traits $+1$ assigned to a given 
species. The fact that $n$ is the only species feature that determines
the equilibrium concentration is consequence of the symmetric solution 
$m^\mu = m ~\forall \mu$ of the saddle-point equations. 
Since $\zeta_0 \geq \zeta_1 \geq \ldots
\geq \zeta_p$  the economically more relevant species, i.e., those
characterized by the set of traits
$\xi^\mu = +1 ~\forall \mu$, are the first ones to go extinct when, say, the cooperation
pressure $u$ decreases. 
The condition for the coexistence
of all  species in the ecosystem is then $\zeta_p > 0$  that reduces to
\begin{equation}\label{coex}
p < 1 + \frac{1}{a} + \frac{2u}{a \left ( 1 - a \right )} .
\end{equation}
We note that the species characterized by $n < p/2$ never die out.
A comparison of these analytical results with the numerical solution
of Eq. (\ref{eq_rep0}) for $N=2000$
and a single instance of the interaction matrix is presented in Fig. \ref{fig:1} where
we show  the cumulative distribution  $C(x)$ for $p=5$. For the sake of illustration, 
the corresponding  probability distribution $P(x) \equiv dC/dx$ is also shown 
in the inset.

\begin{figure}[tbc]
\centerline{\epsfig{file=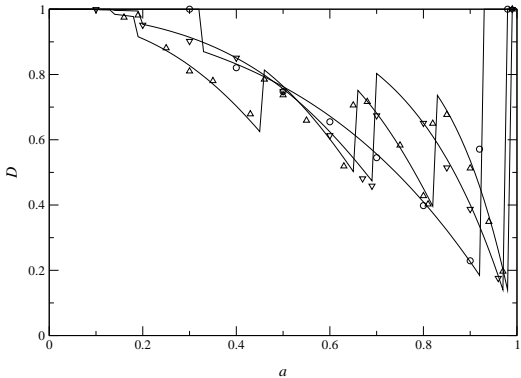,angle=270,width=0.49\textwidth}}
\par
\caption{Fraction of surviving species or diversity D as function of the bias parameter
$a$ for $u=0.1$ and $p=5 ~(\bigcirc )$, $10  ~(\bigtriangledown)$, and
$15 ~(\bigtriangleup)$. The solid curves are the theoretical predictions
and the symbols
are the results of the numerical solution of the replicator equations.}
\label{fig:2}
\end{figure}

The situation of maximum cooperation occurs for $a=0$ since $\zeta_n = 1 ~\forall n$
and any particular set of $p$ traits contributes with the same fraction, $2^{-p}$,
to the final ecosystem composition. However, this symmetry is broken for $a >0$.
In fact, as the result of
the excess of traits $\xi^\mu = +1$, many different species carry the same
or nearly  the same set of traits and so their 
interactions have a strongly competitive character, i.e. $J_{ij}
\approx p/N > 0$,  explaining thus the depletion of the concentration of species with
$n > p/2$. On the other hand, the advantage of species characterized by 
an excess of traits $\xi^\mu = -1$  is twofold. First, their relative
rarenesses  imply that there are few competing species and, second, they have abundance of
cooperative partners among the species with $n > p/2$. These two factors explain
the dominance, in the sense of occuring at larger concentrations and hence
of possessing greater stability, of 
the species characterized by $n < p/2$. A particularly extreme situation is
observed in the  limits $u \to 0$ and $a \to 1$, when  all species with $n > p/2$ 
die out. These findings can be given an obvious ecological interpretation,
namely,  species that are too similar and, consequently, compete for essentially
the same resources are proner to extinction. However, the situation is
not so simple in the case  of a  nonzero $u$. For instance,
in the limit $a \to 1$ the explosive growth of species with $n < p/2$ 
can be controlled even by a vanishingly small self-interaction parameter, resulting  
in the coexistence of all species  [see Eq. (\ref{coex})]. 
Fig. \ref{fig:2} illustrates  the complex dependence of the species 
diversity $D = 1 - C(0)$ on the control parameters $a$ and $p$ for $u=0.1$. 
The results for $u=0$ are very similar, except in
the close neighborhood of $a=1$ where $D \to 0$, implying  that there are  a
finite number of surviving species only. 

These results indicate 
that, when complementarity is the sole mechanism determining the species interactions,
the assemblage of synthetic ecosystems aiming at 
the exploitation of some particular traits (as in monocultures, for instance)
may, in the long term, be disastrous to the economically relevant species. However, 
if there is in addition  some external global pressure for cooperation, then the
best strategy for the long-term survival of those species is to guarantee that they
are massively present in the initial assemblage. Even so, the more stable species are
always those of less economic value.

Although diversity has been manipulated as an independent variable in most of the
numerical studies of model ecosystems as well as in  small scale  experiments,
it is becoming evident that,
in large scales, species diversity itself is a dynamical variable that adjusts
freely to changes in environmental conditions \cite{Loreau}. The understanding of more
complex ecological systems calls then for a more holistic approach.
Though relying on the
unrealistic, technical assumption of symmetry of the interactions between species,
equilibrium statistical mechanics provides an useful (if not the only)
analytic framework to tackle this difficult issue. The 
elusive character of diversity is
apparent from  our results (see, e.g., Fig. \ref{fig:2}) since, despite  the simplicity of 
our model ecosystem, its complex dependence on the control parameters 
of the model precludes a simple description in terms of a single parameter.
This situation is  reminiscent of a recent impass in ecology, 
triggered by the finding of a positive correlation between diversity and productivity in
experiments on randomly assembled communities, whereas in nature the most
productive ecosystems are those characterized by low species diversity \cite{Loreau}.
These antagonistic conclusions may well be due to the attempt to describe 
productivity solely in terms of diversity, while a complete description would require
the knowledge of other, probably uncontrolled, quantities.

\bigskip

The work of J.F.F was supported in part by CNPq and FAPESP, Project No. 99/09644-9.
V.M.O. was supported by FAPESP.

\end{document}